\documentclass[12pt]{article}
\usepackage{amsfonts}
\usepackage{url}

\newtheorem{theorem}{Theorem}

\def\dy{\mathrm{d}y}

\def\dt{\mathrm{d}t}
\def\KL{\mathrm{KL}}

\def\d{\mathrm{d}}

\def\dprob{\mathrm{d}p}
\def\dx{{\mathrm{d}x}}
\def\dtheta{{\mathrm{d}\theta}}
\def\dlambda{{\mathrm{d}\lambda}}
\def\dmu{{\mathrm{d}\mu}}
\def\dsigma{{\mathrm{d}\sigma}}
\def\ds{{\mathrm{d}s}}

\title{Cram\'er-Rao Lower Bound and\\ Information Geometry\footnote{To appear in {\em Connected at Infinity II: On the work of Indian mathematicians} (R. Bhatia and C.S. Rajan, Eds.), special volume of {\em Texts and Readings In Mathematics} (TRIM), Hindustan Book Agency, 2013. \protect\url{http://www.hindbook.com/trims.php}}}
\author{Frank Nielsen\\ Sony Computer Science Laboratories Inc., Japan\\
\'Ecole Polytechnique, LIX, France}
\date{}

\begin{document}
\maketitle

\section{Introduction and historical background}

This article focuses on an important piece of work of the world renowned Indian
statistician, Calyampudi Radhakrishna Rao. In 1945, C. R. Rao (25 years old then) published a pathbreaking paper \cite{rao-1945}, which had a profound impact on subsequent statistical research. Roughly speaking, Rao obtained a lower bound to the variance of an estimator. The importance of this work can be gauged, for instance, by the fact that it has been reprinted in the volume {\em Breakthroughs in Statistics: Foundations and Basic Theory} \cite{StatBreakthroughs:1993}. There have been two major impacts of this work: 

\begin{itemize}

\item First, it answers a fundamental question statisticians have always been interested in, namely, how good can a statistical estimator be? Is there a fundamental limit when estimating statistical parameters?

\item Second, it opens up a novel paradigm by introducing differential geometric modeling ideas to the field of Statistics. In recent years, this contribution has led to the birth of a flourishing field of {\it Information Geometry} \cite{informationgeometry-2000}. 

\end{itemize}

It is interesting to note that H. Cram\'er \cite{Cramer-1946} (1893-1985) also dealt with the same problem in his classic book {\it Mathematical Methods of Statistics}, published in 1946, more or less at the same time Rao's work was published. The result is widely acknowledged nowadays as the Cram\'er-Rao lower bound (CRLB). The lower bound was also reported independently\footnote{The author thanks F. Barbaresco for bringing the historical references to his attention.} in the work of M. Fr\'echet \cite{Frechet-1945} (uniparameter case) and G. Darmois \cite{Darmois-1945} (multi-parameter case). The Fr\'echet-Darmois work were both published in French, somewhat limiting its international scientific exposure. Thus the lower bound is also sometimes called the Cram\'er-Rao-Fr\'echet-Darmois lower bound.

This review article is organized as follows:
Section~\ref{sec:twokeys} introduces the two fundamental contributions in C. R. Rao's paper: 
\begin{itemize}
\item  The Cram\'er-Rao lower bound (CRLB), and 
\item  The Fisher-Rao Riemannian geometry.
\end{itemize}
Section~\ref{sec:state} concisely explains how information geometry has since evolved into a full-fledged discipline. Finally, Section~\ref{sec:conclusion} concludes this review by discussing further perspectives of information geometry and hinting at the future challenges. 

\section{Two key contributions to Statistics\label{sec:twokeys}}
To begin with, we describe the two key contributions of Rao \cite{rao-1945}, namely a lower bound to the variance of an estimator and Rao's Riemannian information geometry.

\subsection{Rao's lower bound for statistical estimators} 
For a fixed integer $n\geq 2$, let $\{X_1, ..., X_n\}$ be a random sample of size $n$ on a random variable $X$ which has a probability density function (pdf) (or, probability mass function (pmf)) $p(x)$. Suppose the unknown distribution $p(x)$ belongs to a parameterized family $\mathcal{F}$ of distributions 
$$
\mathcal{F}=\{p_\theta(x)\ |\ \theta\in\Theta \},
$$ 
where $\theta$ is a parameter vector belonging to the parameter space $\Theta$. For example, $\mathcal{F}$ can be chosen as the family $\mathcal{F}_{\mathrm{Gaussian}}$ of all normal distributions with parameters $\theta=(\mu,\sigma)$ (with $\theta\in\Theta=\mathbb{R}\times \mathbb{R}^+$):

The unknown distribution $p(x)=p_{\theta^*}(x)\in\mathcal{F}$ is identified by a unique parameter $\theta^*\in\Theta$. One of the major problems in Statistics is to build an ``estimator" of $\theta^*$ on the basis of the sample observations $\{X_1,\ldots ,X_n\}$.           

There are various estimation procedures available in the literature, e.g., the method of moments and the method of maximum likelihood; for a more comprehensive account on estimation theory, see e.g., \cite{PointEstimation-1998}. From a given sample of fixed size $n$, one can get several estimators of the same parameter. A natural question then is: which estimator should one use and how their performance compare to each other. This is related precisely with C. R. Rao's  first contribution in his seminal paper~\cite{rao-1945}. Rao addresses the following question:

\noindent What is the {\em accuracy attainable} in the estimation of statistical parameters?

Before proceeding further, it is important to make some observations on the notion of {\em likelihood}, introduced by Sir R. A. Fisher \cite{Fisher-1922}. Let $\{X_1,\ldots ,X_n\}$ be a random vector with 
pdf (or, pmf) $p_{\theta}(x_1,\ldots ,x_n),\;\theta \in \Theta$, where for $1\leq i\leq n$, $x_i$ is a realization of $X_i$. The function 
$$L(\theta;x_1,\ldots ,x_n)=p_{\theta}(x_1,\ldots ,x_n),$$ considered as a function of $\theta$, is called the likelihood function. If $X_1,\ldots ,X_n$ are independent and identically distributed random variables with pdf (or, pmf) $p_{\theta}(x)$ (for instance, if $X_1,\ldots ,X_n$ is a random sample from $p_{\theta}(x)$), the likelihood function is
$$L(\theta;x_1,\ldots ,x_n)=\prod_{i=1}^np_{\theta}(x_i).$$

The method of maximum likelihood estimation consists of choosing an estimator of $\theta$, say $\hat{\theta}$ that maximizes $L(\theta;x_1,\ldots ,x_n)$. If such a $\hat{\theta}$ exists, we call it a {\em maximum likelihood estimator} (MLE) of $\theta$. Maximizing the likelihood function is mathematically equivalent to maximizing the log-likelihood function since the logarithm function is a strictly increasing function. The log-likelihood function is usually simpler to optimize. We shall write $l(x,\theta)$ to denote the log-likelihood function with $x=(x_1,\ldots ,x_n)$. Finally, we recall the definition of an {\em unbiased} estimator. Let $\{ p_{\theta},\theta \in \Theta\}$ be a set of probability distribution functions. An estimator $T$ is said to be an unbiased estimator of $\theta$ if the expectation of $T$, $$E_{\theta}(T)=\theta,\;\;\mbox{for all}\;\theta \in \Theta.$$

Consider probability distributions with pdf (or, pmf) satisfying the following {\em regularity conditions}:

\begin{itemize}
\item The support $\{x \ |\ p_\theta(x)>0 \}$ is identical for all distributions (and thus does not depend on $\theta$),
\item $\int p_\theta(x)\dx$ can be differentiated under the integral sign with respect to $\theta$,
\item The gradient $\nabla_\theta p_\theta(x)$ exists.
\end{itemize}
We are now ready to state C. R. Rao's fundamental limit of estimators.

\subsubsection{Rao's lower bound: Single parameter case}
Let us first consider the case of uni-parameter distributions like Poisson distributions with mean parameter $\lambda$. These families are also called order-$1$ families of probabilities.
The C. R. Rao lower bound in the case of uni-parameter distributions can be stated now. 

\begin{theorem}[Rao lower bound (RLB)]
Suppose the regularity conditions stated above hold. Then the variance of any unbiased estimator  $\hat\theta$, based on an independent and identically distributed (IID) random sample of size $n$, is bounded below by $\frac{1}{nI(\theta^*)}$, where $I(\theta)$ denotes the Fisher information in a single observation, defined as
$$
I(\theta) = -E_\theta\left[\frac{\d^2 l(x;\theta)}{\dtheta^2}\right]=\int  - \frac{\d^2 l(x;\theta)}{\dtheta^2} p_\theta(x) \dx.
$$ 
\end{theorem}

As an illustration, consider the family of Poisson distributions with parameter $\theta=\lambda$.
One can check that the regularity conditions hold. 
For a Poisson distribution with parameter $\lambda$, $l(x;\lambda)=-\lambda + \log \frac{\lambda^x}{x!}$ and thus,

\begin{eqnarray*}
l'(x;\lambda) &=& -1+ \frac{x}{\lambda},\\
l''(x;\lambda)&=&-\frac{x}{\lambda^2}.
\end{eqnarray*}

The first derivative is technically called the {\em score function}. It follows that 
\begin{eqnarray*}
I(\lambda) &=& -E_{\lambda} \left[\frac{\d^2 l(x;\lambda)}{\dlambda^2} \right],\\
& = & \frac{1}{\lambda^2} E_\lambda[x]=  \frac{1}{\lambda}
\end{eqnarray*}
since $E[X]=\lambda$ for a random variable $X$ following a Poisson distribution with parameter $\lambda$: $X\sim \mathrm{Poisson}(\lambda)$. 
What the RLB theorem states in plain words is that for any unbiased estimator $\hat\lambda$ based on an IID sample of size $n$ of a Poisson distribution with parameter $\theta^*=\lambda^*$, the variance of $\hat\lambda$ cannot go  below $\frac{1}{nI(\lambda^*)}=\lambda^*/n$.

The Fisher information, defined as the variance of the score,  can be geometrically interpreted as the curvature of the log-likelihood function.
When the curvature is low (log-likelihood curve is almost flat), we may expect some large amount of deviation from the optimal $\theta^*$.
But when the curvature is high (peaky log-likelihood), we rather expect a small amount of deviation from $\theta^*$.

\subsubsection{Rao's lower bound: Multi-parameter case}

For $d$-dimensional multi-parameter\footnote{Multi-parameter distributions can be univariate like the 1D Gaussians $N(\mu,\sigma)$ or multivariate like the Dirichlet distributions or $d$D Gaussians.} distributions, the Fisher information matrix $I(\theta)$ is defined as the symmetric matrix with the following entries~\cite{informationgeometry-2000}: 

\begin{eqnarray}
[I(\theta)]_{ij}  &=& E_{\theta}\left[ \frac{\partial}{\partial\theta_i} \log p_\theta(x) \frac{\partial}{\partial\theta_j} \log p_\theta(x) \right],\\
&=&
\int \left(\frac{\partial}{\partial\theta_i} \log p_\theta(x) \frac{\partial}{\partial\theta_j} \log p_\theta(x)\right) p_\theta(x)\dx.
\end{eqnarray}

Provided certain regularity conditions are met (see~\cite{informationgeometry-2000}, section 2.2), the Fisher information matrix can be written {\em equivalently} as:
$$
[I(\theta)]_{ij}  = -E_{\theta}\left[\frac{\partial^2}{\partial\theta_i\partial\theta_j} \log p_\theta(x) \right],
$$
or as:
$$
[I(\theta)]_{ij}  = 4\int_{x\in\mathcal{X}} \frac{\partial}{\partial\theta_i} \sqrt{p_\theta(x)} \frac{\partial}{\partial\theta_j} \sqrt{p_\theta(x)}\dx.
$$

In the case of multi-parameter distributions, the lower bound on the accuracy of unbiased estimators can be extended using the  L\"owner partial ordering on matrices defined by $A\succeq B \Leftrightarrow A-B\succeq 0$,
where $M\succeq 0$ means $M$ is positive semidefinite \cite{Bhatia:2006} (We similarly write $M\succ 0$ to indicate that $M$ is positive definite).

The Fisher information matrix is always positive semi-definite~\cite{PointEstimation-1998}.
It can be shown that the Fisher information matrix of regular probability distributions is positive definite, and therefore always invertible.
Theorem~1 on the lower bound on the inaccuracy  extends to the multi-parameter setting as follows:

\begin{theorem}[Multi-parameter Rao lower bound (RLB)]
Let $\theta$ be a\linebreak vector-valued parameter. Then for an unbiased estimator $\hat\theta$  of $\theta^*$ based on a IID random sample of $n$ observations, one has 
$V[\hat\theta] \succeq n^{-1}I^{-1}(\theta^*)$,
where $V[\hat\theta]$ now denotes the variance-covariance matrix of $\hat{\theta}$ and $I^{-1}(\theta^*)$ denotes the inverse of the Fisher information matrix evaluated at the optimal parameter $\theta^*$.
\end{theorem}

As an example, consider a IID random sample of size $n$ from a normal population $N(\mu^*,{\sigma^*}^2)$, so that 
$\theta^*=(\mu^*,{\sigma^*}^2)$. One can then verify that the Fisher information matrix  of a normal distribution $N(\mu,{\sigma}^2)$ is given by
$$I(\theta)=\left[
\begin{array}{cc}
\frac{1}{\sigma^2}&0\\
0&\frac{1}{2\sigma^4}
\end{array}
\right].$$
Therefore,
$$
V[\hat\theta] \succeq
n^{-1}I(\theta^*)^{-1}=\left[
\begin{array}{cc}
n^{-1}{\sigma^*}^2&0\\
0&2n^{-1}{\sigma^*}^4
\end{array}
\right].$$

There has been a continuous flow of research along the lines of the CRLB, including the case where the Fisher information matrix is singular (positive semidefinite, e.g. in statistical mixture models). We refer the reader to the book of Watanabe~\cite{singularinfogeo-2009} for a modern algebraic treatment of degeneracies in statistical learning theory.
\subsection{Rao's Riemannian information geometry}
What further makes C. R. Rao's 1945 paper a truly impressive milestone in the development of Statistics is the introduction of differential geometric methods for modeling population spaces using the Fisher information matrix. Let us review the framework that literally opened up the field of information geometry~\cite{informationgeometry-2000}. 

Rao \cite{rao-1945} introduced the notions of the Riemannian Fisher information metric and geodesic distance to the Statisticians. This differential geometrization of Statistics gave birth to what is known now as the field of {\em information geometry}  \cite{informationgeometry-2000}. Although there were already precursor geometric work \cite{Mahalanobis:1936,Bhatta1943,Mahalanobis:1948} linking geometry to statistics by the Indian  community (Professors Mahalanobis and Bhattacharyya), none of them studied the differential concepts and made the connection with the Fisher information matrix. C. R. Rao is again a  pioneer in offering Statisticians the geometric lens.

\subsubsection{The population space}

Consider a family of parametric probability distribution $p_\theta(x)$ with $x\in\mathbb{R}^d$ and $\theta\in\mathbb{R}^D$ denoting the $D$-dimensional parameters of distributions (order of the probability family). The population parameter space is the space
$$
\Theta=\left\{ \theta\in\mathbb{R}^D \Big |\ \int p_\theta(x)\dx=1\right\}.
$$ 
A given distribution $p_\theta(x)$ is interpreted as a corresponding point indexed by $\theta\in\mathbb{R}^D$. $\theta$ also encodes a coordinate system to identify probability models: $\theta\leftrightarrow p_{\theta}(x)$.

Consider now two infinitesimally close points $\theta$ and $\theta+\dtheta$.
Their probability densities differ by their first order differentials: $\dprob(\theta)$.
The distribution of $\dprob$ over all the support aggregates the consequences of replacing $\theta$ by $\theta+\dtheta$. Rao's revolutionary idea was to consider the {\it relative discrepancy} $\frac{\dprob}{p}$ and to take the variance of this difference distribution to define the following quadratic differential form:
\begin{eqnarray*}
\ds^2(\theta) &=& \sum_{i=1}^D \sum_{j=1}^{D} g_{ij}(\theta) \dtheta_i\dtheta_j,\\
&=& (\nabla \theta)^{\top} G(\theta) \nabla\theta,
\end{eqnarray*}
with the matrix entries of $G(\theta)=[g_{ij}(\theta)]$ as
\begin{eqnarray*}
g_{ij}(\theta)=E_\theta\left[ \frac{1}{p(\theta)} \frac{\partial p}{\partial\theta_i} \frac{1}{p(\theta)} \frac{\partial p}{\partial\theta_j}  \right]=g_{ji}(\theta).
\end{eqnarray*}
In differential geometry, we often use the symbol $\partial_i$ as a shortcut to $\frac{\partial}{\partial\theta_i}$.

The elements $g_{ij}(\theta)$ form the quadratic differential form defining the elementary length of Riemannian geometry. The matrix $G(\theta)=[g_{ij}(\theta)]\succ 0$ is positive definite and turns out to be equivalent to the {\em Fisher information matrix}: $G(\theta)=I(\theta)$.
The information matrix is invariant to monotonous transformations of the parameter space~\cite{rao-1945} and makes it a good candidate for a Riemannian metric.

We shall discuss later more on the concepts of invariance in statistical manifolds \cite{cencov-1982,MorozovaChentsov:1991}.

In \cite{rao-1945}, Rao proposed a novel versatile notion of statistical distance induced by the Riemannian geometry beyond the traditional Mahalanobis $D$-squared distance \cite{Mahalanobis:1936} and the Bhattacharyya distance \cite{Bhatta1943}. The Mahalanobis $D$-squared distance \cite{Mahalanobis:1936} of a vector $x$ to a group of vectors with covariance matrix $\Sigma$ and mean $\mu$ is defined originally as
$$
D^2_\Sigma(x,\mu)=(x-\mu)^\top \Sigma^{-1} (x-\mu).
$$
The generic Mahalanobis distance $D_M(p,q)=\sqrt{(p-q)^\top M (p-q)}$ (with $M$ positive definite) generalizes the Euclidean distance ($M$ chosen as the identity matrix).

The Bhattacharyya  distance \cite{Bhatta1943} between two distributions indexed by parameters $\theta_1$ and $\theta_2$ is defined by
$$
B(\theta_1,\theta_2)=-\log \int_{x\in\mathcal{X}} \sqrt{ p_{\theta_1}(x) p_{\theta_2}(x)}\dx.
$$
Although the Mahalanobis distance $D_M$ is a metric (satisfying the triangle inequality and symmetry), the symmetric Bhattacharyya  distance fails the triangle inequality. Nevertheless, it can be used to define the Hellinger metric distance $H$ whose square is related the Bhattacharyya distance as follows

\begin{equation}\label{eq:HB}
H^2(\theta_1,\theta_2) = \frac{1}{2} \int (\sqrt{ p_{\theta_1}(x)} -\sqrt{p_{\theta_2}(x)})^2 \dx = 1-e^{-B(\theta_1,\theta_2)} \leq 1
\end{equation}

\subsubsection{Rao's distance: Riemannian distance between two populations}

Let $P_1$ and $P_2$ be two  points of the population space corresponding to the distributions with respective parameters $\theta_1$ and $\theta_2$.
In Riemannian geometry, the geodesics are the {\it shortest paths}. 
For example, the geodesics on the sphere are the arcs of great circles.
The statistical distance between the two populations is defined by integrating the infinitesimal element lengths $\ds$ along the geodesic linking $P_1$ and $P_2$. 
Equipped with the Fisher information matrix tensor $I(\theta)$, the {\it Rao distance} $D(\cdot,\cdot)$ between two distributions on a statistical manifold  can be calculated from the geodesic length as follows:
$$
D(p_{\theta_1}(x),p_{\theta_2}(x)) = \min_{\stackrel{\theta(t)}{\theta(0)=\theta_1,\theta(1)=\theta_2}}
\int_0^1  \left(\sqrt{ (\nabla \theta)^\top I(\theta)  \nabla \theta }\right)\dt
$$
Therefore  we need to calculate explicitly the geodesic linking $p_{\theta_1}(x)$ to $p_{\theta_2}(x)$ to compute Rao's distance.
This is done by solving the following second order ordinary differential equation (ODE) \cite{informationgeometry-2000}:
$$
g_{ki} \ddot{\theta_i} + \Gamma_{k,ij} \dot{\theta_i} \dot{\theta_j} = 0,
$$
where Einstein summation \cite{informationgeometry-2000} convention has been used to simplify the mathematical writing by removing the leading sum symbols.
The coefficients $\Gamma_{k,ij}$ are the Christoffel symbols of the first kind  defined by:
$$
\Gamma_{k,ij}= \frac{1}{2}\left( 
\frac{\partial g_{ik}}{\partial\theta_j}
+
\frac{\partial g_{kj}}{\partial\theta_i}
-
\frac{\partial g_{ij}}{\partial\theta_k}
\right).
$$
For a parametric statistical manifold with $D$ parameters, there are $D^3$ Christoffel symbols.
In practice, it is difficult to explicitly compute  the geodesics of the Fisher-Rao geometry of arbitrary models, and one needs to perform a gradient
descent to find a local solution for the geodesics \cite{entropicmetric-shapegeodesic-2006}.
This is a drawback of the Rao's distance as it has to be checked manually whether the integral admits a closed-form expression or not.

To give an example of the Rao distance, consider the smooth manifold of univariate normal distributions, indexed by the $\theta=(\mu,\sigma)$ coordinate system.
The Fisher information matrix is
\begin{equation}
I(\theta)=\left[
\begin{array}{cc}
\frac{1}{\sigma^2} & 0\\
0 & \frac{2}{\sigma^2}
\end{array}
\right] \succ 0.
\end{equation}
The infinitesimal element length is:
\begin{eqnarray*}
\ds^2 &=& (\nabla\theta)^{\top} I(\theta) \nabla\theta,\\
 &=& \frac{\dmu^2}{\sigma^2}+\frac{2\dsigma^2}{\sigma^2}.
\end{eqnarray*} 

After the minimization of the path length integral, the Rao distance between two normal distributions \cite{rao-1945,Atkinson-1981} $\theta_1=(\mu_1,\sigma_1)$ and $\theta_2=(\mu_2,\sigma_2)$ is given by:
$$
D(\theta_1,\theta_2) = \left\{
\begin{array}{ll} 
\sqrt{2} \log \frac{\sigma_2}{\sigma_1} & \mbox{if $\mu_1=\mu_2$},\\
 \frac{|\mu_1-\mu_2|}{\sigma} & \mbox{if $\sigma_1=\sigma_2=\sigma$},\\
  \sqrt{2} \log \frac{\tan \frac{a_1}{2}}{\tan \frac{a_2}{2}} & \mbox{otherwise.}
  \end{array}
  \right.
$$
where $a_1=\arcsin \frac{\sigma_1}{b_{12}}$, $a_2=\arcsin \frac{\sigma_2}{b_{12}}$ and 
$$
b_{12}=\sigma_1^2+ \frac{(\mu_1-\mu_2)^2-2(\sigma_2^2-\sigma_1^2)}{8(\mu_1-\mu_2)^2}.
$$
For univariate normal distributions, Rao's distance amounts to computing the hyperbolic distance for $\mathbb{H}(\frac{1}{\sqrt{2}})$, see \cite{Lovric-2000}.

Statistical distances play a key role in tests of significance and classification \cite{rao-1945}.
Rao's distance is a metric  since it is a Riemannian geodesic distance, and thus satisfies the triangle inequality. Rao's Riemannian geometric modeling of the population space is now commonly called the Fisher-Rao geometry \cite{FisherRao-IJCV-2011}.
One drawback of the Fisher-Rao geometry is the computer tractability of dealing with Riemannian geodesics.
The following section concisely reviews the field of information geometry.
\section{A brief overview of information geometry\label{sec:state}}

Since the seminal work of Rao \cite{informationgeometry-2000} in 1945, the interplay of differential geometry with statistics has further strengthened and developed into a new discipline called {\it information geometry} with a few dedicated monographs \cite{DifferentialGeometryStats-1987,MurrayRice-1993,KassVos-1997,informationgeometry-2000,hessianstructure-2007,infogeo-2008}.
It has been proved by Chentsov and published in his Russian monograph in 1972 (translated in English in 1982 by the AMS \cite{cencov-1982}) that the Fisher information matrix is the {\it only} invariant Riemannian metric for statistical manifolds (up to some scalar factor).
Furthermore, Chentsov \cite{cencov-1982} proved that there exists a family of connections, termed the $\alpha$-connections, that ensures statistical invariance.
\subsection{Statistical invariance and $f$-divergences}

A divergence is basically a smooth statistical distance that may not be symmetric nor satisfy the triangle inequality.
We denote by $D(p:q)$ the divergence from distribution $p(x)$ to distribution $q(x)$, where the ``:'' notation emphasizes the fact that this dissimilarity measure may not be symmetric: $D(p:q)\not = D(q:p)$.

It has been proved that the only statistical invariant divergences \cite{informationgeometry-2000,fdivergenceinvariance-2010} are the Ali-Silvey-Csisz\'ar $f$-divergences $D_f$ \cite{AliSilvey-1966,Csiszar-1967} that are defined for a functional convex generator $f$ satisfying $f(1)=f' (1)=0$ and $f''(1)=1$ by:
$$
D_f (p:q) = \int_{x\in\mathcal{X}} p(x) f\left(\frac{q(x)}{p(x)}\right)\dx.
$$
Indeed, under an invertible mapping function (with $\dim \mathcal{X}=\dim\mathcal{Y}=d$): 
\begin{eqnarray*}
m: &\mathcal{X}& \rightarrow \mathcal{Y} \\
     &x& \mapsto y=m(x)
\end{eqnarray*}
a probability density $p(x)$ is converted into another density $q(y)$ such that: 
$$
p(x)\dx=q(y)\dy,\qquad \dy=|M(x)|\dx,
$$
where $|M(x)|$ denotes the determinant of the Jacobian matrix \cite{informationgeometry-2000} of the transformation $m$ (i.e., the partial derivatives):
$$
M(x) = \left[
\begin{array}{lll}
\frac{\partial y_1}{\partial x_1} & \dots & \frac{\partial y_1}{\partial x_d}\\
\vdots & \ddots & \vdots \\
\frac{\partial y_d}{\partial x_1} & \dots & \frac{\partial y_d}{\partial x_d}
\end{array}
\right].
$$
It follows that 
$$
q(y)=q(m(x)) =p(x)|M(x)|^{-1}.
$$ 
For any two densities $p_1$ and $p_2$, we have the $f$-divergence on the transformed densities $q_1$ and $q_2$ that can be rewritten mathematically as
\begin{eqnarray*}
D_f(q_1:q_2) &=& \int_{y\in\mathcal{Y}} q_1(y) f\left(\frac{q_2(y)}{q_1(y)}\right)\dy,\\
&=&  \int_{x\in\mathcal{X}} p_1(x) |M(x)|^{-1} f\left(\frac{p_2(x)}{p_1(x)}\right) |M(x)|\dx,\\
&=& D_f(p_1:p_2).
\end{eqnarray*}
Furthermore, the $f$-divergences are the only divergences satisfying  the remarkable data-processing theorem \cite{Dataprocessingtheorem-1997} that characterizes the property of information monotonicity \cite{AmariCsiszarBregman-2009}.
Consider discrete distributions on an alphabet $\mathcal{X}$ of $d$ letters. 
For any partition $\mathcal{B}=\mathcal{X}_1\cup ... \mathcal{X}_b$ of $\mathcal{X}$ that  merge alphabet letters into $b\leq d$ bins, we have
$$
0\leq D_f(\bar p_1: \bar p_2) \leq D_f(p_1:p_2 ),
$$
where $\bar p_1$ and $\bar p_2$ are the discrete distribution induced by the partition $\mathcal{B}$ on $\mathcal{X}$.
That is, we loose discrimination power by coarse-graining the support of the distributions.

The most fundamental $f$-divergence is the Kullback-Leibler divergence \cite{ct-1991} obtained for the generator $f(x)=x\log x$:
$$
\KL(p:q) = \int p(x)\log \frac{p(x)}{q(x)}\dx.
$$
The Kullback-Leibler divergence between two distributions $p(x)$ and $q(x)$ is equal to the cross-entropy $H^\times(p:q)$ minus the Shannon entropy $H(p)$:
\begin{eqnarray*}
\KL(p:q) &=& \int p(x)\log \frac{p(x)}{q(x)}\dx,\\
&=& H^\times(p:q)-H(p).
\end{eqnarray*}
with
\begin{eqnarray*}
H^\times(p:q) &=& \int -p(x)\log q(x)\dx,\\
H(p) &=&  \int -p(x)\log p(x)\dx =H^\times(p:p). 
\end{eqnarray*}
The Kullback-Leibler divergence $\KL(\tilde p:p)$ \cite{ct-1991} can be interpreted  as the distance between the estimated distribution $\tilde p$ (from the samples) and the true hidden distribution $p$.

\subsection{Information and sufficiency}

In general, statistical invariance is characterized under Markov morphisms \cite{MorozovaChentsov:1991,fdivergenceinvariance-2010} (also called sufficient stochastic kernels \cite{fdivergenceinvariance-2010}) that generalizes the deterministic transformations $y=m(x)$.
Loosely speaking, a geometric parametric statistical manifold $\mathcal{F}=\{p_\theta(x) | \theta\in\Theta\}$ equipped with a $f$-divergence  must also provide invariance by:

\begin{description}

\item[Non-singular parameter reparameterization.] That is, if we choose a different coordinate system, say $\theta'=f(\theta)$ for an invertible transformation $f$, it should not impact the intrinsic distance between the underlying distributions. 
For example, whether we parametrize the Gaussian manifold by $\theta=(\mu,\sigma)$ or by
$\theta'=(\mu^3,\sigma^2)$, it  should preserve the distance. 

\item[Sufficient statistic.] When making statistical inference, we use statistics $T: \mathbb{R}^d\rightarrow \Theta\subseteq\mathbb{R}^D$ (e.g., the mean statistic $T_n(X)=\frac{1}{n}\sum_{i=1}^n X_i$ is used for estimating the parameter $\mu$ of Gaussians). 
In statistics, the concept of {\it sufficiency} was introduced by Fisher \cite{Fisher-1922}:

``... the statistic chosen should summarize the whole of the relevant information supplied by the sample. ''

Mathematically, the fact that all information should be aggregated inside the sufficient statistic is written as 
$$
\Pr(x|t,\theta)=\Pr(x|t).
$$

It is not surprising that all statistical information of a parametric distribution with $D$ parameters can be recovered from a set of $D$ statistics. For example, the univariate Gaussian with $d=\dim\mathcal{X}=1$ and $D=\dim\Theta=2$ (for parameters $\theta=(\mu,\sigma)$) is recovered from the  mean and variance statistics. 
A sufficient statistic is a set of statistics that compress   information without loss for statistical inference.
\end{description}

\subsection{Sufficiency and exponential families}

The  distributions admitting finite sufficient statistics are called the exponential families \cite{Koopman-1936,BrownExpFam:1986,informationgeometry-2000}, and have their probability density or mass functions canonically rewritten as
$$
p_\theta(x) = \exp (\theta^\top t(x)-F(\theta)+k(x) ),
$$
where $k(x)$ is an auxiliary carrier measure, $t(x):\mathbb{R}^d \rightarrow \mathbb{R}^D$ is the sufficient statistics, and $F: \mathbb{R}^D\rightarrow \mathbb{R}$ a strictly convex and differentiable function, called the cumulant function or the log normalizer since,
$$
F(\theta) = \log \int_{x\in\mathcal{X}} \exp (\theta^\top t(x)+k(x) ) \dx.
$$
See \cite{informationgeometry-2000} for canonical decompositions of usual distributions (Gaussian, multinomial, etc.).
The space $\Theta$ for which the log-integrals converge is called the natural parameter space.

For example, 
\begin{itemize}
\item Poisson distributions are univariate exponential distributions of order $1$ 
(with $\mathcal{X}=\mathbb{N}^*=\{0, 1, 2, 3, ...\}$ and $\dim\Theta=1$) with associated probability mass function:
$$
\frac{\lambda^k e^{-\lambda}}{k!},
$$ 
for $k\in \mathbb{N}^*$.

The canonical exponential family decomposition yields
\begin{itemize}
	\item $t(x)=x$:  the sufficient statistic,
	\item $\theta=\log\lambda$: the natural parameter,
	\item $F(\theta)=\exp\theta$: the cumulant function,
	\item $k(x)=-\log x!$:  the carrier measure. 
\end{itemize}

\item Univariate Gaussian distributions are  distributions of order $2$ (with $\mathcal{X}=\mathbb{R}$, $\dim\mathcal{X}=1$ and $\dim\Theta=2$),  characterized by two parameters $\theta=(\mu,\sigma)$ with associated density:
$$
\frac{1}{\sigma\sqrt{2\pi}} e^{-\frac{1}{2}\left(\frac{x-\mu}{\sigma}\right)^2},
$$
for $x\in\mathbb{R}$.

The canonical exponential family decomposition yields:
\begin{itemize}
	\item $t(x)=(x,x^2)$: the sufficient statistic,
	\item $\theta=(\theta_1,\theta_2)=(\frac{\mu}{\sigma^2},-\frac{1}{2\sigma^2})$: the natural parameters,
	\item $F(\theta)=-\frac{\theta_1^2}{4\theta_2} + \frac{1}{2} \log \left( -\frac{\pi}{\theta_2} \right)$: the cumulant function,
	\item $k(x)=0$: the carrier measure. 
\end{itemize}
\end{itemize}

Exponential families provide a generic framework in Statistics, and are universal density approximators \cite{ExpfamUniversal:2004}.
That is, any distribution can be arbitrarily approximated closely by an exponential family.
An exponential family is defined by the functions $t(\cdot)$ and $k(\cdot)$, and a member of it by a natural parameter $\theta$.
The cumulant function $F$ is evaluated by the log-Laplace transform. 

To illustrate the generic behavior of exponential families in Statistics~\cite{BrownExpFam:1986}, let us consider the maximum likelihood estimator for a distribution  belonging to the  exponential family. We have the MLE $\hat\theta$:
$$
\hat\theta = (\nabla F)^{-1}\left(\sum_{i=1}^n \frac{1}{n} t(x_i) \right),
$$ 
where $(\nabla F)^{-1}$ denotes the reciprocal gradient of $F$: $(\nabla F)^{-1}\circ \nabla F=\nabla F\circ (\nabla F)^{-1}=\mathrm{Id}$, the identity function on $\mathbb{R}^D$.
The Fisher information matrix of an exponential family is
$$
I(\theta) = \nabla^2 F(\theta)\succ 0,
$$
the Hessian of the log-normalizer, always  positive-definite since $F$ is strictly convex.

\subsection{Dual Bregman divergences and $\alpha$-Divergences}

The Kullback-Leibler divergence between two distributions belonging to the same exponential families can be expressed equivalently as a Bregman divergence on the swapped natural parameters defined for the cumulant function $F$ of the exponential family:
\begin{eqnarray*}
\KL(p_{F,\theta_1}(x) : p_{F,\theta_2}(x) ) &=& B_F(\theta_2:\theta_1),\\
&=& F(\theta_2)-F(\theta_1)-(\theta_2-\theta_1)^\top\nabla F(\theta_1)
\end{eqnarray*}
As mentioned earlier, the ``:'' notation emphasizes that the distance is not a metric: It does not satisfy the symmetry nor the triangle inequality in general.
Divergence $B_F$ is called a Bregman divergence \cite{Bregman67}, and is the canonical distances of dually flat spaces \cite{informationgeometry-2000}.
This  Kullback-Leibler divergence on densities $\leftrightarrow$ divergence on parameters relies on the dual canonical parameterization of exponential families \cite{BrownExpFam:1986}. 
A  random variable $X\sim p_{F,\theta}(x)$, whose distribution belongs to an exponential family, can be dually indexed by its expectation parameter $\eta$ such that
$$
\eta=E[t(X)]=\int_{x\in\mathcal{X}} x e^{\theta^\top t(x)-F(\theta)+k(x)} \dx= \nabla F(\theta).
$$ 
For example, the $\eta$-parameterization of Poisson distribution is: $\eta=\nabla F(\theta)=e^\theta=\lambda=E[X]$ (since $t(x)=x$).

In fact, the Legendre-Fenchel convex duality is at the heart of information geometry:
Any strictly convex and differentiable function $F$ admits a dual convex conjugate $F^*$ such that:
$$
F^*(\eta) =  \max_{\theta\in\Theta} \theta^\top \eta-F(\theta).
$$
The maximum is attained for $\eta=\nabla F(\theta)$ and is unique since $F(\theta)$ is strictly convex ($\nabla^2 F(\theta)\succ 0$). 
It follows that $\theta=\nabla F^{-1}(\eta)$, where $\nabla F^{-1}$ denotes the functional inverse gradient.
This implies that:
$$
F^*(\eta) =  \eta^\top (\nabla F)^{-1}(\eta) - F((\nabla F)^{-1}(\eta)).
$$
The Legendre transformation is also called slope transformation since it maps $\theta\rightarrow\eta=\nabla F(\theta)$, where $\nabla F(\theta)$ is the gradient at $\theta$, visualized as the slope of the support tangent plane of $F$ at $\theta$.
The transformation is an involution for strictly convex and differentiable functions: $(F^*)^*=F$.
It follows that gradient of convex conjugates are reciprocal to each other: $\nabla F^*=(\nabla F)^{-1}$.
Legendre duality induces dual  coordinate systems:
\begin{eqnarray*}
\eta=\nabla F(\theta),\\
\theta=\nabla F^*(\eta).
\end{eqnarray*}
Furthermore, those dual coordinate systems are orthogonal to each other since,
$$
\nabla^2 F(\theta) \nabla^2 F^*(\eta) = \mathrm{Id},
$$
the identity matrix.

The Bregman divergence can also be rewritten in a canonical mixed coordinate form $C_F$ or in the $\theta$- or $\eta$-coordinate systems as
\begin{eqnarray*}
B_F(\theta_2:\theta_1) &=& F(\theta_2) + F^*(\eta_1) -\theta_2^\top \eta_1 = C_F(\theta_2,\eta_1) = C_{F^*}(\eta_1,\theta_2),\\
 &=& B_{F^*}(\eta_1:\eta_2).\\
\end{eqnarray*}

Another use of the Legendre duality is to interpret the log-density of an exponential family as a dual Bregman divergence \cite{bregmankmeans-2005}:
$$
\log p_{F,t,k,\theta}(x) = -B_{F^*}(t(x):\eta)+F^*(t(x))+k(x),
$$ 
with $\eta=\nabla F(\theta)$ and $\theta=\nabla F^*(\eta)$.

The Kullback-Leibler divergence (a $f$-divergence) is a particular divergence belonging to the $1$-parameter family of divergences, called $\alpha$-divergences (see~\cite{informationgeometry-2000}, p. 57). The $\alpha$-divergences are defined for $\alpha\not=\pm 1$ as 
$$
 D_\alpha(p:q) = \frac{4}{1-\alpha^2}\left( 1- \int p(x)^{\frac{1-\alpha}{2}}q(x)^{\frac{1+\alpha}{2}}\dx \right).
 $$
It follows that $D_\alpha(q:p)=D_{-\alpha}(p:q)$, and in the limit case, we have: 
$$
D_{-1}(p:q)=\KL(p:q)=\int p(x)\log \frac{p(x)}{q(x)}\dx.
$$

Divergence $D_1$ is also called the reverse Kullback-Leibler divergence, and divergence $D_0$ is four times the squared Hellinger distance mentioned earlier in eq.~\ref{eq:HB}

$$
D_0(p:q) = D_0(q:p) = 4 \left(1-\int \sqrt{p(x)}\sqrt{q(x)}\dx\right) = 4 H^2(p,q).
$$

In the sequel, we denote by $D$ the divergence $D_{-1}$ corresponding to the Kullback-Leibler divergence.

\subsection{Exponential geodesics and mixture geodesics}

Information geometry as further pioneered by Amari \cite{informationgeometry-2000} considers dual affine geometries introduced by a pair of connections: the $\alpha$-connection and $-\alpha$-connection instead of taking the Levi-Civita connection induced by the Fisher information Riemmanian metric of Rao. 
The $\pm 1$-connections give rise to dually flat spaces \cite{informationgeometry-2000} equipped with the Kullback-Leibler divergence \cite{ct-1991}.
The case of $\alpha=-1$ denotes the mixture family, and the exponential family is obtained for $\alpha=1$.
We omit technical details in this expository paper, but refer the reader to the monograph \cite{informationgeometry-2000} for details.

For our purpose, let us say that the geodesics are defined not anymore as shortest path lengths (like in the metric case of the Fisher-Rao geometry) but rather as curves that ensures the parallel transport of vectors \cite{informationgeometry-2000}.
This defines the notion of ``straightness'' of lines. 
Riemannian geodesics satisfy both the straightness property and the minimum length requirements.
Introducing dual connections, we do not have anymore distances interpreted as curve lengths, but the geodesics defined by the notion of straightness only.

In information geometry, we have dual geodesics that are expressed for the exponential family (induced by a convex function $F$) in the dual affine coordinate systems $\theta/\eta$ for $\alpha=\pm 1$ as:
\begin{eqnarray*}
\gamma_{12} &:& L(\theta_1,\theta_2) = \{\theta=(1-\lambda)\theta_1+\lambda\theta_2\ |\ \lambda\in [0,1]\} ,\\
\gamma^*_{12} &:& L^*(\eta_1,\eta_2) = \{\eta=(1-\lambda)\eta_1+\lambda\eta_2\ |\ \lambda\in [0,1]\}.
\end{eqnarray*}
Furthermore, there is a Pythagorean theorem that allows one to define information-theoretic projections \cite{informationgeometry-2000}.
Consider three points $p, q$ and $r$ such that $\gamma_{pq}$ is the $\theta$-geodesic linking $p$ to $q$, and $\gamma^*_{qr}$ is the $\eta$-geodesic linking $q$ to $r$.
The geodesics are orthogonal at the intersection point $q$ if and only if the Pythagorean relation is satisfied:
$$
D(p:r) = D(p:q) + D(q:r).
$$
In fact, a more general triangle relation (extending the law of cosines) exists:
$$
D(p:q) + D(q:r) - D(p:r) = (\theta(p)-\theta(q))^\top (\eta(r)-\eta(q)).
$$
Note that the $\theta$-geodesic $\gamma_{pq}$ and $\eta$-geodesic $\gamma^*_{qr}$ are orthogonal  with respect to the inner product $G(q)$ defined at $q$ (with $G(q)=I(q)$ being the Fisher information matrix at $q$).
Two vectors $u$ and $v$ in the tangent place $T_q$ at $q$ are said to be orthogonal if and only if their inner product equals zero:
$$
u\perp_q v \Leftrightarrow u^\top I(q) v =0.
$$
Observe that in any tangent plane $T_x$ of the manifold, the inner product induces a squared Mahalanobis distance:
$$
D_x(p,q) = (p-q)^\top I(x) (p-q).
$$
Since $I(x)\succ 0$ is positive definite, we can apply Cholesky decomposition on the Fisher information matrix $I(x)=L(x)L^\top(x)$, 
where $L(x)$ is a lower triangular matrix with strictly positive diagonal entries.

By mapping the points $p$ to $L(p)^\top$ in the tangent space $T_p$, the squared Mahalanobis amounts to computing the squared Euclidean distance $D_E(p,q)=\| p-q\|^2$ in the tangent planes:
\begin{eqnarray*}
D_x(p,q) &=& (p-q)^\top I(x) (p-q),\\
& =& (p-q)^\top L(x) L^\top(x) (p-q),\\
&=& D_E(L^\top(x)p, L^\top(x)q).
\end{eqnarray*}
It follows that after applying the ``Cholesky transformation'' of objects into the tangent planes, we can solve geometric problems in tangent planes as one usually does in the Euclidean geometry. 

Information geometry of dually flat spaces thus extend the traditional self-dual Euclidean geometry, obtained for the convex function $F(x)=\frac{1}{2} x^\top x$ (and corresponding to the statistical manifold of isotropic Gaussians). 

\section{Conclusion and perspectives\label{sec:conclusion}}

Rao' s paper \cite{rao-1945} has been instrumental for the development of modern statistics. 
In this masterpiece, Rao introduced what is now commonly known as the Cram\'er-Rao lower bound (CRLB) and the Fisher-Rao geometry. Both the contributions are related to the Fisher information, a concept due to Sir R. A. Fisher, the father of mathematical statistics \cite{Fisher-1922} that introduced the concepts of consistency, efficiency and sufficiency of estimators.
This paper is undoubtably recognized as the cornerstone for introducing differential geometric methods in Statistics. This seminal work has inspired many researchers and has evolved into the field of information geometry \cite{informationgeometry-2000}.
Geometry is originally the science of Earth measurements.
But geometry is also the science of invariance as advocated  by Felix Klein Erlang's program, the science of intrinsic measurement analysis. 
This expository paper has presented the two key contributions of C. R. Rao in his 1945 foundational paper, and briefly presented information geometry without the burden of differential geometry (e.g., vector fields, tensors, and connections).
Information geometry has now ramified far beyond its initial statistical scope, and is further expanding prolifically in many different new horizons.
To illustrate the versatility of information geometry, let us mention a few research areas:  

\begin{itemize}
\item Fisher-Rao Riemannian geometry \cite{FisherRao-IJCV-2011},
\item Amari's dual connection information geometry \cite{informationgeometry-2000},
\item Infinite-dimensional exponential families and Orlicz spaces \cite{CenaPistone:2007},
\item Finsler information geometry \cite{FinslerIG-2006},
\item Optimal transport geometry \cite{GeometryOptimalTransport:1996},
\item Symplectic geometry, K\"ahler manifolds and Siegel domains \cite{SymplecticKahlerGeometry-2009},
\item Geometry of proper scoring rules \cite{ProperScoringRule-2007},
\item Quantum information geometry \cite{grasselli-2001-4}.
\end{itemize}

Geometry with its own specialized language, where words like distances, balls, geodesics, angles, orthogonal projections, etc., provides ``thinking tools'' (affordances) to manipulate non-trivial mathematical objects and notions.
The richness of geometric concepts in information geometry helps one to  reinterpret, extend or design novel algorithms and data-structures by enhancing creativity.
For example, the traditional expectation-maximization (EM) algorithm \cite{em-1977} often used in Statistics has been reinterpreted and further extended using the framework of information-theoretic alternative projections \cite{IG-EM:1995}.
In machine learning, the famous boosting technique that learns a strong classifier by combining linearly weak weighted classifiers has been revisited \cite{UBoost-2004} under the framework of information geometry.
Another striking example, is the study of the geometry of dependence and Gaussianity for Independent Component Analysis \cite{Cardoso-2003}.

\end{document}